\documentclass[prl,aps,showpacs,preprintnumbers,amsmath,amssymb]{revtex4}
\usepackage{graphicx}
\usepackage{amsmath}
\usepackage{amsfonts}
\usepackage{amssymb}
\providecommand{\U}[1]{\protect\rule{.1in}{.1in}}
\def\be{\begin{equation}}
\def\ee{\end{equation}}
\def\bea{\begin{eqnarray}}
\def\eea{\end{eqnarray}}
\begin{document}

\title{Could Fermion Masses Play a Role in the Stabilization of the Dilaton in
Cosmology? }
\author{Alejandro Cabo$^{*,**,***}$ and Robert Brandenberger$^{*}$}

\affiliation{$^{*}$ Department of Physics, McGill University, 3600
rue Universit\'e, Montr\'eal,
QC, H3A 2T8, Canada} \affiliation{$^{**}%
$ Perimeter Institute for Theoretical Physics, 31 Caroline Street,
Waterloo,
ON, N2L 2Y5, Canada} \affiliation{$^{***}%
$Group of Theoretical Physics, Instituto de Cibernetica,
Matematica y Fisica, Habana, Cuba}

\date{Dec.  28 \ 2008 }

\begin{abstract}
We study the possibility that the Dilaton is stabilized by the contribution of
fermion masses to its  effective potential. We consider the Dilaton gravity
action in four dimensions to which we add a mass term for a Dirac fermion.
Such an action describes the interaction of the Dilaton with the fermions in
the Yang-Mills sector of the coupled supergravity/super-Yang-Mills action
which emerges as the low energy effective action of superstring theory after
the extra spatial dimensions have been fixed. The Dilaton couples to the
Fermion mass term via the usual exponential factor of this field which
multiplies the non-kinetic terms of the matter Lagrangian, if we work in the
Einstein frame. In the kinetic part of the Fermion action in the Einstein
frame the Dilaton does not enter. Such masses can be generated in several
ways: they can arise as a consequence of flux about internal spatial
dimensions, they may arise as thermal fermion masses in a quasi-static phase
in the early universe, and they will arise after the breaking of supersymmetry
at late times. The vacuum contribution to the potential for the Dilaton is
evaluated up to two loops. The result shows a minimum which could stabilize
the Dilaton for reasonable ranges of parameter values. 
\end{abstract}
\maketitle

\section{Introduction}

Superstring theory predicts the existence of the Dilaton, a scalar field
partner of the graviton. Dilaton excitations appear in the mass spectrum of a
superstring at the same level as the graviton fluctuations (see e.g.
\cite{GSW} for a textbook discussion). Consistent backgrounds for superstring
theory must include the Dilaton field in addition to the metric, and the
action for such a background is that of Dilaton gravity (see e.g.
\cite{Ven,TV}).

At the tree level, the Dilaton is a free massless scalar field with a very
specific coupling to the matter sector of the theory. Phenomenologically, the
existence of the Dilaton leads to two sets of problems. Since the Dilaton is
coupled to matter, then a rolling Dilaton leads to time-dependent coupling
constants. To avoid this problem, the Dilaton should be constant at the
current stage of cosmological evolution. This is not sufficient, however:
unless the Dilaton has a potential yielding a large mass, then it would lead
to a ``Fifth force". Fifth force constraints lead to a lower bound on the mass
of the Dilaton today which is of the order $m < 10^{-12} \mathrm{GeV}$
\cite{bound} (but see \cite{Polyakov} for an attempt to make a running Dilaton
consistent with late time cosmology).

The question of Dilaton stabilization at various stages of the evolution of
the universe has therefore been a subject of considerable interest in recent
years. In fact, the Dilaton is just one of the various scalar fields which
follow from superstring theory in the low-energy limit. Other scalar fields
describe the sizes and shapes of the extra spatial dimensions predicted by
superstring theory. Collectively, these fields are called ``moduli fields". In
the context of recent work on moduli stabilization in Type IIB superstring
theory, it has been realized that certain moduli fields - including the
Dilaton - can be stabilized by the addition of fluxes about the compact
spatial dimensions \cite{DS,GKP}. Non-perturbative effects such as gaugino
condensation \cite{gaugino} have also been used to stabilize the Dilaton, in
particular in the context of heterotic superstring theory \cite{heterotic} and
string gas cosmology \cite{Danos}.

Since the issue of Dilaton stabilization is a problem for late time cosmology,
it is of interest to explore possible mechanisms to stabilize the Dilaton
which do not depend on special assumptions about extra dimensions.

Another motivation to look for a different Dilaton stabilization mechanism
comes from String Gas Cosmology (SGC). String Gas Cosmology \cite{BV} (see
e.g. \cite{SGCrevs} for recent reviews) is a model of early universe cosmology
which makes use of new degrees of freedom and new symmetries of string theory,
coupling these ingredients to a classical background model containing gravity
plus the Dilaton. The universe is assumed to start as a compact space filled
with a gas of strings. Since there is a maximal temperature for a gas of
closed strings, the early stage of cosmological evolution in SGC will be
described by a phase of almost constant temperature, the ``Hagedorn phase''.
SGC leads to the possibility of a non-singular cosmology. As has recently been
observed, thermal fluctuations in a gas of closed strings in the Hagedorn
phase can lead to a scale-invariant spectrum of cosmological fluctuations
\cite{NBV,BNPV2}, with a slight blue tilt for gravitational waves forming a
distinctive prediction of the model \cite{BNPV1}. However, in order that this
scenario work, the Dilaton needs to be fixed during the Hagedorn phase. Thus,
in SGC we would like the Dilaton to be fixed both at very early and at very
late times (it need not be fixed in intermediate phases).

The issue of clarifying the way in which the Dilaton field is stabilized in
cosmology is thus an important question for particle physics today. The
universal coupling of the Dilaton to the matter sector of the Lagrangian not
only leads to the conclusion that a moving Dilaton will produce time-dependent
coupling constants, but it also leads to the possibility that quantum effects
due to the interaction of the Dilaton with matter might generate interesting
contributions to the effective potential of the Dilaton. This is the question
we will explore in this paper.

We will focus on periods in cosmology when the extra spatial dimensions of
superstring theory have already been stabilized. Thus, our analysis will be in
the context of field theory in four-dimensional space-time. We will focus on
the interaction of the Dilaton with fermions which have a mass. Such masses
can be produced by fluxes about internal manifolds. In the application of our
work to late time cosmology, we will consider the fermion masses generated
after supersymmetry breaking. In another possible application to early
universe cosmology, we can consider thermal masses for the fermions. We will
be thus discussing the possibility that the appearance of fermion masses could
furnish a mechanism for the stabilization of the Dilaton and also for the
generation for its mass.

For this purpose we start from a simple form for the Dilaton gravity action to
which a massive Dirac fermion term is added \cite{elizalde}. The action is
studied in the Einstein frame in which the kinetic terms for the fermions have
no coupling to the Dilaton field. The coupling of the fermion mass terms to
the Dilaton involves the universal exponential factor of Dilaton gravity
\cite{Ven,TV}. Such an action describes the interaction of Super-Yang-Mills
fermions with the Dilaton in the low energy effective action of superstring
theory, as discussed in the following section. We then evaluate the effective
potential as a function of the Dilaton field up to a two loop correction in a
small Dilaton field limit, assuming a fixed value of the cosmological scale
factor. The interesting outcome is that due to the presence of logarithms in
the loop expansion, the Dilaton field appears in quadratic powers multiplying
the exponential factors of itself, leading to the possibility that minima of
the potential exist. This effect occurs for a wide range of the parameters.

The outline of this paper is as follows: In Section 2, we present the action
which will be studied in the rest of the paper and show how it arises in the
low energy limit of superstring theory. In Section
3, the effective action is employed to write down and evaluate the one and two
loop contributions to the effective potential.  The results are discussed in 
the final section. In an appendix we review the definition
of the zero temperature generating functional and effective action and the
background field method for computing these.
 
\section{Dilaton Gravity with a Massive Fermion}

In this paper we will consider an action of a Dilaton coupled to a massive
fermion of the following simple form:
\begin{equation}
S \, = \, \int dx\sqrt{-g(x)}\left(  -\frac{1}{2}g^{\mu\nu}(x)\partial_{\mu
}\varphi(x)\partial_{\nu}\varphi(x)+\overline{\Psi}(x)(i\frac{g^{\mu\nu}
\gamma_{\mu}\overleftrightarrow{\partial}_{\nu}}{2}-\exp(\alpha\text{ }
\varphi)m)\Psi(x)\right)  \, .\label{action}
\end{equation}
This action describes the coupling between the Dilaton $\varphi$ and the
Fermionic part of super-Yang-Mills matter in the Einstein frame, as we will
show in the following. The form of introducing the derivative
$\overleftrightarrow{\partial}=\overrightarrow{\partial}-\overleftarrow
{\partial}$ assures the reality of the Lagrangian in curved space
\cite{elizalde}. Note that the kinetic term contains no Dilaton-dependent
exponential factor, whereas the mass term does. This is demanded if we start
from the low-energy action for superstring theory, as we show below. The Dirac
matrices satisfy the usual commutation relations $\left\{  \gamma_{\mu}
,\gamma_{\nu}\right\}  = 2g_{\mu\nu}(x)$, where $g_{\mu\nu}$ is the space-time
metric, $g^{\mu\nu}$ is its inverse, and $g$ is its determinant.

Our Lagrangian (\ref{action}) describes the coupling between the Dilaton
$\varphi$ and matter fermions $\Psi$ resulting from the low energy effective
action of superstring theory. The starting point is the Lagrangian for
supergravity in ten space-time dimensions written in the Einstein frame:
\begin{equation}
L_{SG} \, = \, {\frac{1 }{{2 \kappa^{2}}}} R - {\frac{3 }{4}} e^{- (3/2)
\varphi} H_{MNP}^{2} - {\frac{ 9 }{{16 \kappa^{2}}}} (\partial_{M}
\varphi)^{2} \, .
\end{equation}
In the above, $R$ is the Ricci scalar, $\kappa^{2}$ is the gravitational
constant in ten dimensions, $\varphi$ is the Dilaton, $H_{MNP}$ is the three
form which is the curl of the fundamental string theoretical two form $B_{MN}$
, and the capital Latin indices run over all space-time dimensions (see
Chapter 13 of \cite{GSW}). We have neglected the contribution from the
gravitini and the dilatini.

To this Lagrangian, we add the Lagrangian of the super-Yang-Mills matter
sector which is
\begin{equation}
L_{SYM} \, = \, - {\frac{1 }{4}} e^{- (3/4) \varphi} F_{MN}^{a} F^{MNa} -
{\frac{1 }{2}} {\bar\chi}^{a} \Gamma^{M} (D_{M} \chi)^{a} + {\frac{1 }{{16}}}
\sqrt{2} e^{- (3/4) \varphi} {\bar\chi}^{a} \Gamma^{MNP} H_{MNP} \chi^{a} \,
,\label{SYMLag}
\end{equation}
where $F_{MN}^{a}$ is the field strength of the gauge field of matter (the
index $a$ denotes the index in group space), $\chi$ is the matter sector
fermion field and the $\Gamma^{M}$ are constant Dirac algebra matrices out of
which the $\Gamma$ matrices with more indices are constructed by taking
totally antisymmetrized products. Once again, we have neglected the
contribution from the gravitini and the dilatini. In the presence of matter,
the three form field $H_{MNP}$ gets a contribution from the Chern-Simons three
form. Since we will in the following be neglecting the contribution of the
gauge fields, the Chern-Simons term will not appear, either.

{F}rom (\ref{SYMLag}) it follows that a non-vanishing three form flux about
internal spatial dimensions can generate a mass term for the fermions $\chi$.
If we fix gravity, then the action for the remaining fields (the Dilaton and
the fermion $\chi$) takes on the form of our toy model (\ref{action}) with
$\alpha= - 3/4$.

\section{Effective Potential}

In the following, we will study the one and two loop approximations of the
above effective action at zero temperature. The background field method
of computing the effective action is reviewed in the Appendix. In this
paper, we are mainly interested in local quantities such as the effective
potential, and hence it is justified to work in the flat space-time
approximation. Note that working in curved space-time would entail very
nontrivial complications.

We will thus be working with the Minkowski metric $g_{\mu\nu}=\eta_{\mu\nu}$
and can view this approximation as working in comoving coordinates and
conformal time $\eta$, the latter being defined in terms of the cosmological
time $t$ via
\begin{equation}
d\eta\,=\,a(t)^{-1}dt,
\end{equation}
(where $a(t)$ is the cosmological scale factor), and neglecting the evolution
of the scale factor. For conformally coupled matter, this procedure is
rigorous. However, in our case we are interested in mass generation, and hence
our procedure is only an approximation (albeit a good one).

Therefore, the generating functional can be written in the form
\begin{align}
Z[j,\overline{\eta},\eta] \,  &  = \, \exp\left\{  i\left[  \int dx \sqrt
{-g}\left(  -\frac{1}{2}g^{\mu\nu}(x)\partial_{\mu}\phi(x)\partial_{\nu}%
\phi(x)\right)  +\int dx\text{{}}j(x)\phi(x)\right]  \right\} \nonumber\\
&  \times\frac{Det\left[  {\LARGE (}i\frac{\gamma^{\mu}\overleftrightarrow
{\partial}_{\mu}}{2}-\exp(\alpha\text{ }\phi)m{\LARGE )}\right]  }{Det\left[
\partial^{2})\right]  ^{\frac{1}{2}}}\label{shifted}\\
&  \times\exp\left[  -i\text{ }\int dx\exp(\alpha\phi)\text{ }m\left(
\exp(\alpha\frac{\delta}{i\text{ }\delta j(x)})-1\right)  \frac{\delta
}{-i\delta\eta(x)}\frac{\delta}{i\delta\overline{\eta}(x)}\right] \nonumber\\
&  \times\exp[-i\int\int dx_{1}dx_{2}\overline{\eta}(x_{1})G(x_{1},x_{2}%
|\phi)\eta(x_{2})]\exp[-i\text{ }\int\int dx_{1}dx_{2}\text{\ }j(x_{1}%
)\frac{D(x_{1},x_{2})}{2}j(x_{2})] \, ,\nonumber
\end{align}
where the metric has been kept in the general form in the tree Dilaton part of
the effective action since replacing it by the Minkowski metric is not
required in terms of possible technical simplifications.

The inverse propagators now have the form they take on in Minkowski
space-time:
\begin{align}
D^{-1}(x_{1},x_{2}) \,  &  = \, \partial_{(1)}^{2}\delta(x_{1}-x_{2}),\\
G^{-1}(x_{1},x_{2}|\phi) \,  &  = \, [i\gamma^{\nu}\partial_{\nu}^{(1)}%
-\exp(\alpha\text{ }\phi)m]\text{ }\delta(x_{1}-x_{2}),
\end{align}
Therefore, the effective action for the Dilaton field with the Fermi fields
set to zero can be written as follows:
\begin{align}
\Gamma\lbrack\phi,0,0] \,  &  \equiv\, \Gamma\lbrack\phi] \, = \, \frac{1}%
{i}\log(Z[j,0,0])-\int dx\text{ }j(x)\phi\label{effective}\\
&  = \, \int dx\sqrt{-g(x)}\left(  -\frac{1}{2}g^{\mu\nu}(x)\partial_{\mu}%
\phi(x)\partial_{\nu}\phi(x)\right) \nonumber\\
&  + \, \frac{1}{i}\log Det{\LARGE [}i\frac{\gamma^{\mu}\overleftrightarrow
{\partial}_{\mu}}{2}-\exp(\alpha\text{ }\phi)m{\LARGE ]}-\frac{1}{2i}\log
Det{\LARGE [}\partial^{2}{\LARGE ]}\nonumber\\
&  + \, \frac{1}{i}\log\left[  \exp\left\{  -i\text{ }\int dx\exp(\alpha
\phi)\text{ }m\left(  \exp(\alpha\frac{\delta}{i\text{ }\delta j(x)}%
)-1\right)  \frac{\delta}{-i\delta\eta(x)}\frac{\delta}{i\delta\overline{\eta
}(x)}\right\}  \times\right. \nonumber\\
&  \left.  \exp{\Large [}-i\int\int dx_{1}dx_{2}\overline{\eta}(x_{1}%
)G(x_{1},x_{2}|\phi)\eta(x_{2}){\Large ]}\exp{\Large [}-i\text{ }\int\int
dx_{1}dx_{2}\text{\ }j(x_{1})\frac{D(x_{1},x_{2})}{2}j(x_{2}){\Large ]}%
\right]  \, .\nonumber
\end{align}
The first term on the right hand side of (\ref{effective}) is the tree level
effective action. We now turn to the evaluation of the one loop contribution.

\subsection{One Loop Effective Action}

Using standard techniques we can write down the one loop correction to the
Effective action associated with vacuum fluctuations. According to the
background field method, the one loop effective action $\Gamma^{(1)}(\phi)$
can be obtained by calculating the Gaussian approximation of the generating
functional about the background field (see e.g. \cite{SC,RMP} for reviews). In
terms of the shifted fields, this corresponds to the Gaussian approximation of
(\ref{shifted}) about the point where the field fluctuations are zero.

Returning to (\ref{shifted}), we see that the exponent of the first term on
the right hand side of the equation is the tree level effective action,
whereas the one loop contribution resulting from the Gaussian integral
approximation is the logarithm of the two determinants appearing in the second
line of (\ref{shifted}). Thus, the sum of the tree and one loop vacuum
effective actions takes the form
\begin{align}
\Gamma^{(0,1)}[\phi] \,  &  = \, \Gamma^{(0)}[\phi]+\Gamma^{(1)}%
[\phi]\nonumber\\
&  = \, \int dx\sqrt{-g(x)}\left(  -\frac{1}{2}g^{\mu\nu}(x)\partial_{\mu}%
\phi(x)\partial_{\nu}\phi(x)\right) \\
&  + \frac{1}{i}\log Det{\LARGE [}i\frac{\gamma^{\mu}\overleftrightarrow
{\partial}_{\mu}}{2}-\exp(\alpha\text{ }\phi)m{\LARGE ]}-\frac{1}{2i}\log
Det{\LARGE [}\partial^{2}{\LARGE ].}\nonumber
\end{align}
The last line is the one loop correction $\Gamma^{(1)}[\phi]$, the first
determinant coming from the fermionic term, the second from the term involving
Dilaton fluctuations.

The functional determinants appearing in $\Gamma^{(1)}[\phi]$ can be evaluated
by going to the momentum space representation. Making use of the identity
$\mathrm{ln}( \mathrm{det}) = \mathrm{Tr}(\mathrm{ln})$ we can write the
fermionic term as
\begin{equation}
\Gamma^{(1)}_{f} \, = \, {\frac{1 }{i}} V^{4} \int{\frac{{dp^{4}} }%
{{(2\pi)^{4}}}} \mathrm{Tr}^{4} \mathrm{ln} \bigl[e^{2 \alpha\phi}m^{2} -
p^{2} \bigr]%
\end{equation}
where the remaining trace runs over the Dirac indices, and the superscripts
$4$ over the volume, the trace and the integration measure indicates the
dimension of space-time. A similar expression holds for the contribution to
the one loop effective potential coming from Dilaton loops (whose
contribution, however, will vanish in the Minimal Subtraction scheme used
below). In order to evaluate this integral, we perform Wick rotation. The
trace over the Dirac indices is trivial and results in a multiplicative factor
of $4$.

The momentum integral diverges. To regularize and renormalize the resulting
expression for the one loop contribution to the effective action, we evaluate
the Euclidean integrals using dimensional regularization. We consider the
integrals in $D$ space-time dimensions. In $D \neq4$ the integral converges
and can be evaluated exactly. The renormalization consists of subtracting the
contribution which gives the pole at $D = 4$. As is known from the
Coleman-Weinberg construction \cite{CW}, we need to introduce a mass scale
$\mu$ to define the subtraction point (``dimensional transmutation"). Given
this mass scale, we can write the D-dimensional volume as
\begin{equation}
V^{D} \, = \, V^{4}\mu^{4-D} \, ,
\end{equation}
and we can re-scale the coupling constant $\alpha$ as
\begin{equation}
\alpha^{D} \, = \, \alpha\mu^{2-\frac{D}{2}} \, .
\end{equation}

After Wick rotation and going to a general dimension, the expression for the
one-loop effective potential has the form
\begin{align}
\Gamma^{(1)}[\phi] \,  &  = \, \frac{1}{i}\log Det{\LARGE [}i\frac{\gamma
^{\mu}\overleftrightarrow{\partial}_{\mu}}{2}-\exp(\alpha^{D}\text{ }
\phi)m{\LARGE ]}-\frac{1}{2i}\log Det{\LARGE [}\partial^{2}{\LARGE ]}%
\nonumber\\
&  = \, 2V^{D}\int\frac{dp^{D}}{(4\pi)^{D}}\log{\LARGE [}p^{2}+\exp
(2\alpha^{D}\text{ }\phi)m^{2}{\LARGE ]}-\frac{1}{2}V^{D}\int\frac{dp^{D}
}{(4\pi)^{D}}\log{\LARGE [}p^{2}{\LARGE ]}\nonumber\\
&  = \, 2V^{D}\frac{\Gamma(1-\frac{D}{2})}{(4\pi)^{\frac{D}{2}}}\frac
{\exp(\alpha^{\text{ }D}D\text{ }\phi)\text{ }m^{D}}{\frac{D}{2}%
}.\label{poten1}
\end{align}
Note that in $D$ space-time dimensions the fields and $\alpha$ have the
following energy dimensions:
\begin{align}
D[\phi] \,  &  = \, \frac{D}{2}-1,\\
D[\Psi] \,  &  = \, \frac{D-1}{2},\\
D[\alpha^{D}\phi] \,  &  = \, 0\rightarrow D[\alpha^{D}] \, = \, 1-\frac{D}
{2}.
\end{align}

After evaluation of the integrals in $D$ space-time dimensions and after
eliminating the divergences at $D=4$ employing the Minimal Subtraction (MS)
scheme, that is, by omitting the pole part in $D-4$ of the result in
(\ref{poten1}), the finite part of the potential becomes (after some
algebra):
\begin{align} \label{oneloop}
V^{(1)}[\phi] \,  & = \,-\frac{\Gamma_{finite}^{(1)}[\phi]}{V} \, = \,
(\frac{m^{2}}{4\pi})^{2}\exp(4\alpha\phi){\Large [}\frac{3}{2} -\gamma
-\log(\frac{m^{2}}{4\pi\mu^{2}}) - 2\text{ }\alpha\text{ }\phi{\Large ] + C}\\
\gamma\,  & = \, 0.57721 \,.\nonumber
\end{align}
An undetermined contribution $C$ to the cosmological constant has been added
in the above expression for possible future applications to phenomenological
questions. Note that the general equation for the effective action always
admits this addition.

As a function of $\phi$, the one-loop potential (\ref{oneloop}) has a local
maximum. Thus, we conclude that at one-loop level, there is no Dilaton stabilization.

\begin{figure}[h]
\includegraphics[width=3in]{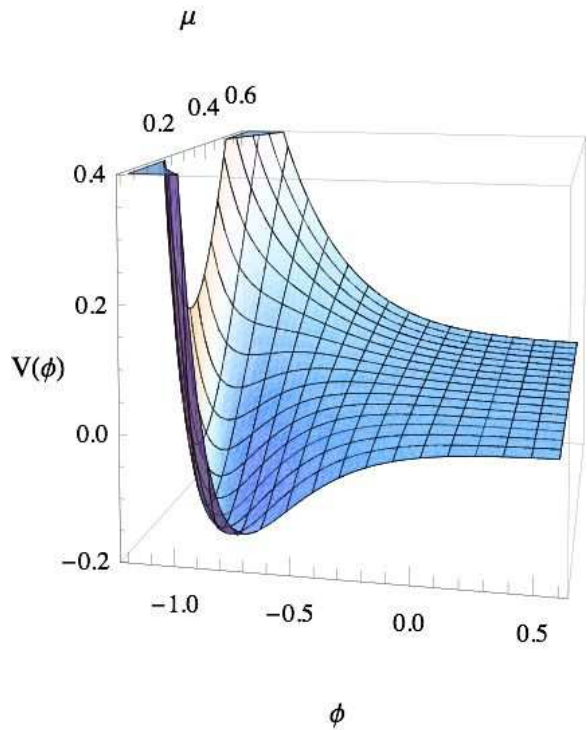}\caption{The sum of the one and two
loop contributions to the effective potential plotted, for fixed values of the
parameters $m=1$, $C=0$ and $\alpha=-1 $, as a function of $\mu$ and the mean
value of the Dilaton field $\phi$. Note the appearance of a minimum of the
potential as a function of the Dilaton field which becomes more pronounced for
smaller values of $\mu$. }%
\label{fig1mod}%
\end{figure}

\subsection{Vacuum two loop correction for small $\varphi$}

Let us consider now the highly non-linear interaction term between the Dilaton
and the fermion field in the action (\ref{action}). We will assume that the
mean value of the Dilaton and of its the quantum fluctuations are small in
order that the functional integral (\ref{gen1}) is not affected by only
retaining the order one term in the Taylor expansion of the $exp(\alpha
\varphi)$ in powers of $\varphi$. Thus, the action of the system becomes
equivalent to the corresponding one in the Yukawa field theory.

In this "small Dilaton" approximation the two loop correction to the effective
action can be written as the \ 1PI\ diagram \ term
\begin{equation}
\Gamma^{(2)}[\phi] \, = \, - \frac{1}{2}\alpha^{2}\exp(2\alpha\phi)m^{2}\int
dx_{1}dx_{2}D(x_{1},x_{2})G^{r_{2}r_{1}}(x_{2},x_{1}|\phi)G^{r_{1}r_{2}}
(x_{1},x_{2}|\phi)
\end{equation}
After expressing the above in terms of momentum integrations by Fourier
transforming the propagators, we obtain:
\begin{align}
\Gamma^{(2)}[\phi] \,  &  =  \, -\frac{1}{2}\alpha^{2}\mu^{4-D}V^{D}%
\exp(2\alpha\phi)m^{2}\int\frac{dp_{1}}{(2\pi)^{D}}\frac{dp_{2}}{(2\pi)^{D}%
}\frac{1}{-(p_{1}-p_{2})^{2}}\nonumber\\
&   \, \times\left(  \frac{1}{\gamma^{\mu}p_{_{1}\mu}-\exp(\alpha\phi)\text{
}m}\right)  ^{r_{2}r_{1}}\left(  \frac{1}{\gamma^{\mu}p_{_{2}\mu}-\exp
(\alpha\phi)m}\right)  ^{r_{1}r_{2}}\nonumber\\
&  =  \, -\frac{1}{2}\alpha^{2}\mu^{4-D}V^{D}\exp(2\alpha\phi)m^{2}\int
\frac{dp_{1}}{(2\pi)^{D}}\frac{dp_{2}}{(2\pi)^{D}}\frac{1}{-(p_{1}-p_{2})^{2}
}\nonumber\\
&   \, \times\frac{4(p_{2}.p_{3}+\exp(2\alpha\phi)\text{ }m^{2})}{(p_{_{1}%
}^{2} -\exp(2\alpha\phi)\text{ }m^{2})(p_{_{2}}^{2}-\exp(2\alpha\phi)\text{
}m^{2} )},
\end{align}
in which the scalar product between the two integration momenta can be
eliminated by employing the identity
\begin{equation}
p_{2}.p_{3}+\exp(2\alpha\phi)\text{ }m^{2}=\exp(2\alpha\phi)\text{ }
m^{2}-\frac{(p_{1}-p_{2})^{2}}{2}+\frac{p_{1}{}^{2}+p_{2}{}^{2}}{2},
\end{equation}
which allows $\Gamma^{(2)}[\phi]$ to be expressed as follows
\begin{align}
\Gamma^{(2)}[\phi] \,  &  =  \, -\alpha^{2}\mu^{4-D}V^{D}\exp(2\alpha
\phi)m^{2} \int\frac{dp_{1}}{(2\pi)^{D}}\frac{dp_{2}}{(2\pi)^{D}}\nonumber\\
&   \, \times\frac{1}{(p_{_{1}}^{2}-\exp(2\alpha\phi)\text{ }m^{2})(p_{_{2}%
}^{2} -\exp(2\alpha\phi)\text{ }m^{2})}\nonumber\\
&   \, -4\alpha^{2}\mu^{4-D}V^{D}\exp(4\alpha\phi)m^{4}\int\frac{dp_{1}}
{(2\pi)^{D}}\frac{dp_{2}}{(2\pi)^{D}}\frac{1}{-(p_{1}-p_{2})^{2}}\\
&   \, \text{ \ \ \ \ \ \ \ \ \ \ \ \ \ \ \ \ \ \ \ \ \ } \times\frac
{1}{(p_{_{1}} ^{2}-\exp(2\alpha\phi)\text{ }m^{2})(p_{_{2}}^{2}-\exp
(2\alpha\phi)\text{ }m^{2})}.\nonumber
\end{align}

The first integral term in the last expression is the square of a one-loop
integral. The second terms is a two loop integral which expression reduces to
a particular class of a master integral which has been evaluated, by example
in Ref. \cite{davidtausk}. After employing the formulae for these integrals
given in Ref. \cite{davidtausk}, the two-loop correction can be written as
follows
\begin{align}
\Gamma^{(2)}[\phi] \,  &  =  \, -\alpha^{2}(m^{\ast2})^{3}V^{4}\frac
{i^{4\epsilon-6}\pi^{4-2\epsilon}\Gamma^{2}(\epsilon-1)}{(2\pi)^{8-4\epsilon}%
}(-\frac{m^{\ast2}}{\mu^{2}})^{-2\epsilon}\nonumber\\
&   \, -\frac{4\alpha^{2}(m^{\ast2})^{3}\pi^{4-2\epsilon}V^{4}}{(2\pi
)^{4(2-\epsilon)}}\frac{\Gamma^{2}(1+\epsilon)}{\epsilon^{2}(1-\epsilon
)(1-2\epsilon)}(\frac{\pi m^{\ast2}}{\mu^{2}})^{-2\epsilon},\nonumber\\
m^{\ast} \,  &  =  \, m\exp(\alpha\text{ }\phi),\text{ \ \ \ \ \ \ \ \ \ }%
\epsilon\, = \, 2-\frac{D}{2}.
\end{align}

The negative of the finite part of $\Gamma^{(2)}[\phi]$ \ in the limit
$\epsilon->0$ gives the two loop contribution \ to the effective potential
\begin{equation}
V^{(2)}[\phi] \, ,= \, -\frac{\Gamma_{finite}^{(2)}[\phi]}{V^{(4)}}.
\end{equation}

The explicit formula for $V^{(2)}$ can be evaluated using Mathematica and is
given below:
\begin{align}
V^{(2)}(\phi) \,  & = \frac{1}{512 \pi^{4}}e^{6 \alpha\phi} \alpha^{2} m^{6}
\bigl[ 4 \bigl( \log\left( \frac{m^{2}}{\mu^{2}}\right)  +2 \alpha\phi\bigr
)\nonumber\\
&  \times\bigl( 3 \left( \log\left( \frac{m^{2}}{\mu^{2}}\right)  + 2
\alpha\phi\right)  - 2 (5+\log(64)+3 \log(\pi)) + 6 \gamma\bigr)\nonumber\\
&  \,\, + 2 \bigl(25 + 8 \log(2) (5 + \log(8)) + \log(\pi) (20 + 6 \log(16
\pi)) \bigr)\nonumber\\
&  \,\, - 8 \gamma\bigl( 5 + \log(64) + 3 \log(\pi) \bigr) + \pi^{2} + 12
\gamma^{2} \bigr]%
\end{align}
We see that the two-loop contribution to the potential leads to a term which
is proportional to
\begin{equation}
(\alpha\phi)^{2} e^{6 \alpha\phi}%
\end{equation}
which changes the character of the potential at small values of $|\phi|$
dramatically. There is now a local minimum of the potential at a small and
negative value of $\phi$ which becomes deeper the smaller $\mu$ is.

The total effective potential up to two loops is given by
\begin{equation}
V[\phi]\,= \, V^{(1)}[\phi]\,+V^{(2)}[\phi] \, .
\end{equation}
A plot of the resulting potential as a function of the Dilaton field and of
the dimensional renormalization parameter $\mu,$ is shown in Figure
\ref{fig1mod} where we have chosen the particular value of the mass $m=1$ and
the negative parameter $\alpha= -1$.

The plot focuses on a region of the regularization parameter $\mu$ which is of
special interest. It is clear that in this zone, the Dilaton field can be
stabilized at a negative value thanks to the appearance of a minimum value of
the potential at small values of $\mu$ relative to $m$, which for simplicity
was taken equal to one for the plot.

\section{Discussion and Conclusions}

In this paper we have computed one and two-loops contributions to the
effective potential of the Dilaton coupled to Dirac fermions in a way
motivated by the low energy effective action of superstring theory. The
divergences in the resulting integrals were regularized and renormalized using
dimensional regularization and Minimal Subtraction. This introduces a new mass
scale $\mu$ into the result. Our result shows that if the ratio of the Fermion
mass $m$ to the new mass scale $\mu$ is sufficiently large, a minimum in the
Dilaton effective potential appears at two-loop order. This leads to a
possible mechanism that could stabilize the Dilaton field. 

Note that since the minimum of the potential is an absolute minimum, the
``overshoot problem" \cite{Brustein} is not an issue. A more important
worry, on the other hand, is the stability of our stabilization mechanism to
higher loop corrections. It is difficult to address this worry in a perturbative
framework.

An immediate application of our result is to the stabilization of the Dilaton
at late times, after the time of supersymmetry breaking. At that time, we
would be making use of the fermions of the Standard Model. Supersymmetry is
broken and so the use of our model Lagrangian is justified.

The application of our results to early universe cosmology are more
speculative. Our model Lagrangian is not supersymmetric: we have neglected the
gravitini and dilatini. In a fully supersymmetric background, we would expect
loops of the fields we have neglected to result in terms cancelling the
contributions to the effective potential which we have computed. However, at
finite density and finite temperature in the early universe supersymmetry is
broken. Thus, one might in fact hope that our one loop potential is also
applicable in early universe cosmology. In this context, one would use the
thermal masses of fermions to stabilize the Dilaton.

In most cosmological background the scale factor changes rapidly with time in
the early universe. This leads to another concern about the applicability of
our analysis to early universe cosmology. However, in String Gas Cosmology, it
is postulated that the early dense phase is a static Hagedorn phase at constant
temperature. In this context, our analysis is justified, hence providing a new
way of stabilizing the Dilaton in this phase of String Gas Cosmology.

Concerning the possible application of our mechanism to Dilaton stabilization
in superstring cosmology, there are further caveats: Not only the Dilaton, but
also the shape and size moduli of the extra dimensions need to be stabilized.
Fluxes which wrap cycles of the extra-dimensional manifold and
non-perturbative techniques such as gaugino condensation are often used for
this purpose (see e.g. \cite{Eva} for an overview). These mechanism can also
generate a potential for the Dilaton which can compete with the potential we
have derived here.

The form of the potential suggests that after solving for the cosmological
evolution of the model, the thermal energy of the Universe could be gradually
transformed in energy of the Dilaton, which then could play the role of a
quintessence field describing Dark Energy. The study of this possibility will
be considered elsewhere.

\bigskip

\begin{acknowledgments}
We are grateful to  Niayesh Afshordi, Keshav Dasgupta, Andrew
Frey, Ghazal Geshnizjani and in particular to Claudia de Rham and
Justin Khoury for very helpful discussions and comments. We wish to
thank Jim Cline for asking a key question which led us to discover
a wrong sign   in one term of the original version of Eq. (\ref{oneloop}%
). The work
of R.B. is supported an NSERC Discovery Grant and by the Canada
Research Chairs program. A.C. would like to deeply acknowledge the
Cosmology Groups of the Department of Physics of McGill University
and of the Perimeter Institute, for the kind invitations to visit
these Centers. His research was  also supported by Proyecto
Nacional de Ciencias Básicas(PNCB) and the Network N-35 of the
Office of External Activities(OEA) of the ICTP.
\end{acknowledgments}

\section{Appendix: Generating Functional and Effective Action}

The generating functional in terms the auxiliary sources $j(x),\eta(x)$ and
${\bar{\eta}(x)}$ for the Dilaton and fermion fields is defined as the
functional integral (see e.g. \cite{SC,RMP})
\begin{equation}
Z[j,\overline{\eta},\eta] \, = \, \int\mathcal{D}\varphi\mathcal{D}
\overline{\Psi} \mathcal{D} \Psi  
\exp\left\{  i[S+\int dx{\large (}j(x)\varphi(x)+\overline
{\eta}(x)\Psi(x)+\overline{\Psi}(x)\eta(x){\large )]}\right\}  .\label{gen1}
\end{equation}

In order to compute the one loop effective action it is beneficial to make use
of the background field method. Thus, we shift the scalar field
\begin{equation}
\varphi(x)\,\rightarrow\,\phi+\varphi^{r}(x),
\end{equation}
where $\phi$ is a classical background Dilaton field which solves the
equations of motion in the absence of fermions. Inserting this expansion into
(\ref{gen1}), $Z[j,\overline{\eta},\eta]$ can be written in the following
form
\begin{align}
Z[j,\overline{\eta},\eta]\, &  =\,\exp\left\{  i\left[  \int dx\sqrt
{-g(x)}\left(  -\frac{1}{2}g^{\mu\nu}(x)\partial_{\mu}\phi(x)\partial_{\nu
}\phi(x)\right)  +\int dx\text{{}}j(x)\phi(x)\right]  \right\}  \nonumber\\
&  \times\exp\left[  -i\text{ }\int dx\sqrt{-g(x)}\exp(\alpha\phi)\text{
}m\left(  \exp(\alpha\frac{\delta}{i\text{ }\delta j(x)})-1\right)
\frac{\delta}{-i\delta\eta(x)}\frac{\delta}{i\delta\overline{\eta}(x)}\right]
\nonumber\\
&  \times\mathcal{D}\varphi_{r}\exp\left\{  i\left[  \int dx\sqrt
{-g(x)}\left(  -\frac{1}{2}g^{\mu\nu}(x)\partial_{\mu}\varphi_{r}%
(x)\partial_{\nu}\varphi_{r}(x)\right)  +\int dx\text{{}}j(x)\varphi
_{r}(x)\right]  \right\}  \nonumber\\
&  \times\int\mathcal{D}\overline{\Psi}\mathcal{D}\Psi\exp\left\{  i\int
dx\left[  \sqrt{-g(x)}{\Huge (}\overline{\Psi}(x){\LARGE (}i\frac{g^{\mu\nu
}\gamma_{\mu}\overleftrightarrow{\partial}_{\nu}}{2}-\exp(\alpha\text{ }%
\phi)m{\LARGE )}\Psi(x){\Huge )}+\right.  \right.  \nonumber\\
&  +\overline{\eta}(x)\Psi(x)+\overline{\Psi}(x)\eta(x){\Huge ]\}.}%
\end{align}

Now, it is possible to evaluate the two Gaussian integrals appearing in the
above expression to obtain
\begin{align}
Z[j,\overline{\eta},\eta] \,  &  = \, \exp\left\{  i\left[  \int
dx\sqrt{-g(x)}\left(  -\frac{1}{2}g^{\mu\nu}(x)\partial_{\mu}\phi
(x)\partial_{\nu}\phi(x)\right)  +\int dx\text{{}}j(x)\phi(x)\right]  \right\}
\nonumber\\
&  \times\frac{Det\left[  \sqrt{-g}{\LARGE (}i\frac{g^{\mu\nu}\gamma_{\mu
}\overleftrightarrow{\partial}_{\nu}}{2}-\exp(\alpha\text{ }\phi
)m{\LARGE )}\right]  }{Det\left[  \partial_{\mu}(\sqrt{-g}g^{\mu\nu
}(x)\partial_{\nu})\right]  ^{\frac{1}{2}}}\\
&  \times\exp\left[  -i\text{ }\int dx\sqrt{-g(x)}\exp(\alpha\phi)\text{
}m\left(  \exp(\alpha\frac{\delta}{i\text{ }\delta j(x)})-1\right)
\frac{\delta}{-i\delta\eta(x)}\frac{\delta}{i\delta\overline{\eta}(x)}\right]
\nonumber\\
&  \times\exp[-i\int\int dx_{1}dx_{2}\overline{\eta}(x_{1})G(x_{1},x_{2}%
|\phi)\eta(x_{2})]\exp[-i\text{ }\int\int dx_{1}dx_{2}\text{\ }j(x_{1}%
)\frac{D(x_{1},x_{2})}{2}j(x_{2})]\nonumber
\end{align}
where the boson and fermion propagators $D(x_{1},x_{2})$ and $G(x_{1} ,
x_{2}|\phi)$ are the inverses of the kernels
\begin{align}
D^{-1}(x_{1},x_{2}) \,  &  = \, \partial_{\mu}^{(1)}(\sqrt{-g(x_{1})}g^{\mu
\nu}(x_{1})\partial_{\nu}^{(1)})\delta(x_{1}-x_{2}),\\
G^{-1}(x_{1},x_{2}|\phi) \,  &  = \, \sqrt{-g(x_{1})}(i\frac{g^{\mu\nu}%
\gamma_{\mu}\overleftrightarrow{\partial}_{\nu}^{(1)}}{2}-\exp(\alpha\text{
}\phi)m)\delta(x_{1}-x_{2}).
\end{align}

The effective action is defined as the Legendre transform of ${\frac{1 }{i}}
\mathrm{ln} Z$, with the canonically conjugate fields being the sources and
the respective mean fields. Thus, we have
\begin{equation}
\Gamma\lbrack\phi,\psi,\overline{\psi}] \,  = \, \frac{1}{i}\log
Z[j,\overline{\eta},\eta]-\int dx{\large (}j(x)\phi(x)+\overline{\eta}
(x)\psi(x)+\overline{\psi}(x)\eta(x){\large )},
\end{equation}
where $\phi(x), \psi(x)$ and $\overline{\psi}(x)$ are the functional derivatives
of the logarithm of the partition function with respect to $i j(x), i \overline{\eta}(x)$
and $-i \eta(x)$, respectively.
Very similar expressions can be written down for the finite temperature
partition function and effective action.

%
%
%
%
%


\begin{thebibliography}{99}
\bibitem{GSW}M. B. Green, J. H. Schwarz and E. Witten, \textit{Superstring
theory} (Cambridge University Press, Cambridge, 1987).

\bibitem{Ven}G.~Veneziano,  ``Scale factor duality for classical and quantum
strings,''  Phys.\ Lett.\ B \textbf{265}, 287 (1991).

\bibitem{TV}A.~A.~Tseytlin and C.~Vafa,  ``Elements of string cosmology,''
Nucl.\ Phys.\ B \textbf{372}, 443 (1992)  [arXiv:hep-th/9109048].

\bibitem{bound}E.~G.~Adelberger, B.~R.~Heckel and A.~E.~Nelson,  ``Tests of
the gravitational inverse-square law,''
Ann.\ Rev.\ Nucl.\ Part.\ Sci.\ \textbf{53}, 77 (2003)
[arXiv:hep-ph/0307284].

\bibitem{Polyakov}T.~Damour and A.~M.~Polyakov,  ``The String Dilaton And A
Least Coupling Principle,''  Nucl.\ Phys.\ B \textbf{423}, 532 (1994)
[arXiv:hep-th/9401069].

\bibitem{DS}K.~Dasgupta, G.~Rajesh and S.~Sethi,  ``M theory, orientifolds and
G-flux,''  JHEP \textbf{9908}, 023 (1999)  [arXiv:hep-th/9908088].

\bibitem{GKP}S.~B.~Giddings, S.~Kachru and J.~Polchinski,  ``Hierarchies from
fluxes in string compactifications,''  Phys.\ Rev.\ D \textbf{66}, 106006
(2002)  [arXiv:hep-th/0105097].

\bibitem{gaugino}S.~Ferrara, L.~Girardello and H.~P.~Nilles,  ``Breakdown Of
Local Supersymmetry Through Gauge Fermion Condensates,''  Phys.\ Lett.\ B
\textbf{125}, 457 (1983);\newline I.~Affleck, M.~Dine and N.~Seiberg,
``Supersymmetry Breaking By Instantons,''  Phys.\ Rev.\ Lett.\ \textbf{51},
1026 (1983);\newline I.~Affleck, M.~Dine and N.~Seiberg,  ``Dynamical
Supersymmetry Breaking In Supersymmetric QCD,''  Nucl.\ Phys.\ B \textbf{241},
493 (1984);\newline I.~Affleck, M.~Dine and N.~Seiberg,  ``Dynamical
Supersymmetry Breaking In Four-Dimensions And Its  Phenomenological
Implications,''  Nucl.\ Phys.\ B \textbf{256}, 557 (1985);\newline
M.~A.~Shifman and A.~I.~Vainshtein,  ``On Gluino Condensation in
Supersymmetric Gauge Theories. SU(N) and O(N)  Groups,''  Nucl.\ Phys.\ B
\textbf{296}, 445 (1988)  [Sov.\ Phys.\ JETP \textbf{66}, 1100 (1987)].

\bibitem{heterotic}M.~Dine, R.~Rohm, N.~Seiberg and E.~Witten,  ``Gluino
Condensation In Superstring Models,''  Phys.\ Lett.\ B \textbf{156}, 55
(1985).

\bibitem{Danos}R.~J.~Danos, A.~R.~Frey and R.~H.~Brandenberger,  ``Stabilizing
moduli with thermal matter and nonperturbative effects,''  
Phys.\ Rev.\  D {\bf 77}, 126009 (2008)
  [arXiv:0802.1557 [hep-th]].

\bibitem{BV}R.~H.~Brandenberger and C.~Vafa,  ``Superstrings In The Early
Universe,''  Nucl.\ Phys.\ B \textbf{316}, 391 (1989).

\bibitem{SGCrevs}R.~H.~Brandenberger,  ``String gas cosmology and structure
formation: A brief review,''  Mod.\ Phys.\ Lett.\ A \textbf{22}, 1875 (2007)
[arXiv:hep-th/0702001];
\newline R.~H.~Brandenberger,  ``Moduli stabilization
in string gas cosmology,''  Prog.\ Theor.\ Phys.\ Suppl.\ \textbf{163}, 358
(2006)  [arXiv:hep-th/0509159];
\newline T.~Battefeld and S.~Watson,  ``String
gas cosmology,''  Rev.\ Mod.\ Phys.\ \textbf{78}, 435 (2006)
[arXiv:hep-th/0510022].

\bibitem{NBV}A.~Nayeri, R.~H.~Brandenberger and C.~Vafa,  
``Producing a scale-invariant spectrum of perturbations in a Hagedorn phase  of string
cosmology,''  
Phys.\ Rev.\ Lett.\  {\bf 97}, 021302 (2006)
  [arXiv:hep-th/0511140];
\newline A.~Nayeri,  ``Inflation free,
stringy generation of scale-invariant cosmological  fluctuations in D = 3 + 1
dimensions,''  arXiv:hep-th/0607073.

\bibitem{BNPV2}R.~H.~Brandenberger, A.~Nayeri, S.~P.~Patil and C.~Vafa,
``String gas cosmology and structure formation,''  Int.\ J.\ Mod.\ Phys.\ A
\textbf{22}, 3621 (2007)  [arXiv:hep-th/0608121].

\bibitem{BNPV1}R.~H.~Brandenberger, A.~Nayeri, S.~P.~Patil and C.~Vafa,
``Tensor modes from a primordial Hagedorn phase of string cosmology,''
Phys.\ Rev.\ Lett.\ \textbf{98}, 231302 (2007)  [arXiv:hep-th/0604126].

\bibitem{elizalde}E.~Elizalde, S.~Naftulin and S.~D.~Odintsov,  ``One-loop
divergence in dilaton gravitation with neutral fermions,''  Phys.\ Rev.\ D
\textbf{49}, 2852 (1994)  [Russ.\ Phys.\ J.\ \textbf{37}, 903
(1994\ IVUFA,37N9,116-121.1994)]  [arXiv:hep-th/9308020].

\bibitem{SC}S.~R.~Coleman,  ``Secret Symmetry: An Introduction To Spontaneous
Symmetry Breakdown And  Gauge Fields,''  Lectures given at 1973 Int. Summer
School of Physics Ettore Majorana,  Erice, Sicily, Jul 6-25, 1973,  Published
in Erice Subnucl.Phys.1973:139

\bibitem{RMP}R. H. Brandenberger,  ``Quantum Field Theory Methods And
Inflationary Universe Models,''  Rev.\ Mod.\ Phys.\ \textbf{57}, 1 (1985).

\bibitem{CW}S.~R.~Coleman and E.~Weinberg,  ``Radiative Corrections As The
Origin Of Spontaneous Symmetry Breaking,''  Phys.\ Rev.\ D \textbf{7}, 1888
(1973).

\bibitem{davidtausk}A.~I.~Davydychev and J.~B.~Tausk,  ``Two-loop self-energy
diagrams with different masses and the momentum  expansion,'' Nucl.\ Phys.\ B
\textbf{397}, 123 (1993).

\bibitem{Brustein}
R.~Brustein and P.~J.~Steinhardt,
  ``Challenges for superstring cosmology,''
  Phys.\ Lett.\  B {\bf 302}, 196 (1993)
  [arXiv:hep-th/9212049].

\bibitem{Eva}E.~Silverstein, ``TASI / PiTP / ISS lectures on moduli and
microphysics,'' arXiv:hep-th/0405068.

%
\end{thebibliography}
\end{document}